\documentclass[manuscript]{acmart}

\AtBeginDocument{%
  \providecommand\BibTeX{{%
    \normalfont B\kern-0.5em{\scshape i\kern-0.25em b}\kern-0.8em\TeX}}}

\setcopyright{acmcopyright}
\copyrightyear{2018}
\acmYear{2018}
\acmDOI{XXXXXXX.XXXXXXX}
\usepackage{algorithm}
\usepackage{listings, xcolor}

\definecolor{verylightgray}{rgb}{.97,.97,.97}

\lstdefinelanguage{Solidity}{
	keywords=[1]{anonymous, assembly, assert, balance, break, call, callcode, case, catch, class, constant, continue, constructor, contract, debugger, default, delegatecall, delete, do, else, emit, event, experimental, export, external, false, finally, for, function, gas, if, implements, import, in, indexed, instanceof, interface, internal, is, length, library, log0, log1, log2, log3, log4, memory, modifier, new, payable, pragma, private, protected, public, pure, push, require, return, returns, revert, selfdestruct, send, solidity, storage, struct, suicide, selfdestruct, super, switch, then, this, throw, transfer, true, try, typeof, using, value, view, while, with, addmod, ecrecover, keccak256, mulmod, ripemd160, sha256, sha3}, % generic keywords including crypto operations
	keywordstyle=[1]\color{blue}\bfseries,
	keywords=[2]{address, bool, byte, bytes, bytes1, bytes2, bytes3, bytes4, bytes5, bytes6, bytes7, bytes8, bytes9, bytes10, bytes11, bytes12, bytes13, bytes14, bytes15, bytes16, bytes17, bytes18, bytes19, bytes20, bytes21, bytes22, bytes23, bytes24, bytes25, bytes26, bytes27, bytes28, bytes29, bytes30, bytes31, bytes32, enum, int, int8, int16, int24, int32, int40, int48, int56, int64, int72, int80, int88, int96, int104, int112, int120, int128, int136, int144, int152, int160, int168, int176, int184, int192, int200, int208, int216, int224, int232, int240, int248, int256, mapping, string, uint, uint8, uint16, uint24, uint32, uint40, uint48, uint56, uint64, uint72, uint80, uint88, uint96, uint104, uint112, uint120, uint128, uint136, uint144, uint152, uint160, uint168, uint176, uint184, uint192, uint200, uint208, uint216, uint224, uint232, uint240, uint248, uint256, var, void, ether, finney, szabo, wei, days, hours, minutes, seconds, weeks, years},	% types; money and time units
	keywordstyle=[2]\color{teal}\bfseries,
	keywords=[3]{block, blockhash, coinbase, difficulty, gaslimit, number, timestamp, msg, data, gas, sender, sig, value, now, tx, gasprice, origin},	% environment variables
	keywordstyle=[3]\color{violet}\bfseries,
	identifierstyle=\color{black},
	sensitive=false,
	comment=[l]{//},
	morecomment=[s]{/*}{*/},
	commentstyle=\color{gray}\ttfamily,
	stringstyle=\color{red}\ttfamily,
	morestring=[b]',
	morestring=[b]"
}
\usepackage{balance}
\usepackage{subfig}

\begin{document}

\title{Mining Domain Models in Ethereum DApps using Code Cloning}

\author{Noama Fatima Samreen}

\email{noama.samreen@ryerson.ca}
\affiliation{%
  \institution{Ryerson University}
  \streetaddress{Department of Computer Science}
  \city{Toronto}
  \state{Ontario}
  \country{Canada}
}

\author{Manar H. Alalfi}

\email{manar.alalfi@ryerson.ca}
\affiliation{%
  \institution{Ryerson University}
  \streetaddress{Department of Computer Science}
  \city{Toronto}
  \state{Ontario}
  \country{Canada}
}

\begin{abstract}
This paper discusses and demonstrates the use of near-miss clone detection to support the characterization of domain models of smart contracts for each of the popular domains in which smart contracts are being rapidly adopted. In this paper, we leverage the code clone detection techniques to detect similarities in functions of the smart contracts deployed onto the Ethereum blockchain network. We analyze the clusters of code clones and the semantics of the code fragments in the clusters in an attempt to categorize them and discover the structural models of the patterns in code clones.
\end{abstract}



\maketitle

\section{Introduction}\label{sec1}

Model-driven engineering (MDE) make use of structural and behavioral models that can be used to help a novice programmer of smart contracts understand the new features provided by the blockchain-specific programming languages such as Solidity \cite{Solidity}. These models provide a way of abstracting technical and language-dependent functionalities thereby increasing the structural and behavioral understanding of a new technology such as the blockchain technology. Given the stigma around the complexity of development of decentralised applications (DApps) using smart contracts, model-driven development could enable the widespread adoption of the blockchain technology as software developers of other technologies could use this as a starting point of their blockchain oriented software engineering process. MDE is a strategic framework that is used to realize advanced solutions by considering different aspects and stakeholders involved in a domain. By viewing different models we can achieve separation of concerns with a higher level of abstraction and reduces the complexity of dealing with DApps specification.
%The Architecture Driven Modernization (ADM) initiative was released by the Object  Management  Group (OMG) in 2003\cite{OMG}. ADM was launched for implementing the principles of MDE technology in software modernization. The main goal of implementing ADM is to produce a set of standard Meta Models. The modernization of software systems using ADM involves understanding and evolving existing software to maintain their business value.
%The development of an MDE framework for Ethereum smart contracts using the ADM process could be realised as follows:
%\begin{itemize}
%    \item Extraction of Knowledge-Discovery Meta Models (KDM) from deployed Ethereum smart contracts
%    \item Generation of artefacts involved in a modernization process using the KDM Meta Models
%\end{itemize} 

Code clone detection in software systems has many applications. For example, it can be used to identify repeated similar code fragments with the aim of developing libraries and standards. It can also be used to enable bug fixes, updates and changes to a software system \cite{androidCodeClone}.

In this paper, we leverage code clone detection technique by using Nicad \cite{NiCad} to detect similarities in functions of the smart contracts deployed onto the Ethereum blockchain network. We analyse the clusters of code clones and the semantics of the code fragments in the clusters in an attempt to categorise them and discover the structural models of the patterns in code clones. These structural models of patterns produced by code clone detection technique could then be used as conceptual models in defining low-level semantic Meta Models. These structural domain models are discovered by reverse engineering of the largest code clone clusters in the source code files of smart contracts deployed on the Ethereum blockchain network. 

Our aim in this research is to identify patterns of code reuse through code clone detection technique to build models on top of the existing code patterns in Ethereum smart contracts. 
%We are working towards an ultimate goal of developing an MDE framework (see Figure \ref{figure:initial}) for Ethereum smart contracts development that follows the approach defined by Cosentino et al. \cite{cosentino} to extract business rules from a Java application using Model-Driven Reverse Engineering (MDRE). Cosentino et al. describe the first step towards achieving their goal is to extract the domain model from the Java application. Similarly, we aim at extracting domain models from Ethereum smart contracts, however and unlike Cosentino et al. approach, we use code clone detection and clustering to ease the process of characterizing the business rules and recovering the target domain models.
We approach this study by investigating the extent of code clones in Ethereum smart contracts, categorizing these clone clusters and producing structural domain models from these clusters to enable code reuse at the modelling level. We highlight that our research leverages an existing code clone detection tool, NiCad that operates at the source code level in order to detect clones of type 1, 2, 2-c, 3 and 3-c.
In this paper, we address the following research questions:  %Therefore, following their approach, we perform code clone detection in Ethereum smart contracts to address the following research questions regarding the code clone clusters and their semantics to deduce domain models: 

\textbf{RQ1:} How can we identify and analyse the code clones present in main-net verified smart contracts that are associated with DApps?

 \textbf{RQ2:} Can the clone clusters of smart contracts be categorised into domains depending on their functionality? 
 
 \textbf{RQ3:} Can the categorised clone clusters of smart contracts be reverse engineered to produce structural domain models?
 
\begin{figure*}[t!]
\centerline{\includegraphics[width=0.9\textwidth]{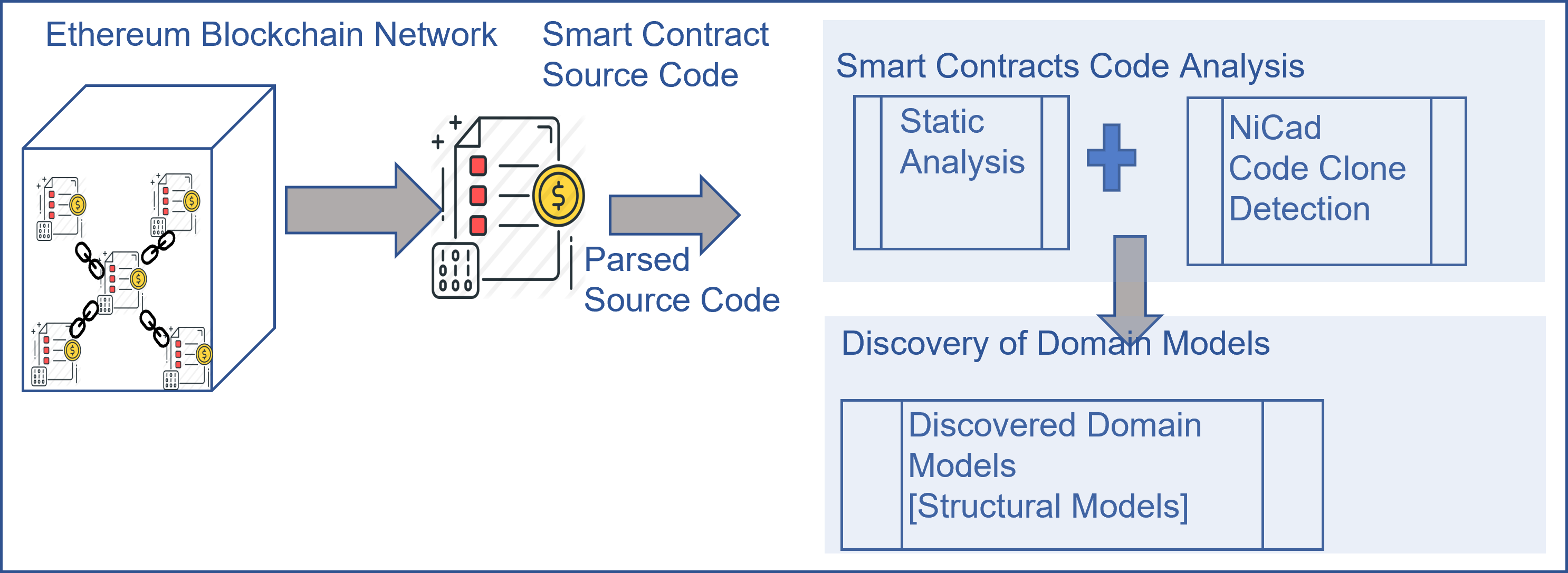}}
\caption{Extraction of Domain Models Workflow}
%\vspace{-0.5 cm}
\label{figure:initial}
\end{figure*}

Our workflow for discovering structural domain models from Ethereum smart contracts consists of following steps: 
\begin{enumerate}
    \item Smart contracts code usage analysis: Identification of highly used code patterns by leveraging code clone detection technique. As BT and smart contracts are just being adopted in various industries, we initiate our models discovery process by analyzing the extent and characterization of code clones in already deployed smart contracts.
    \item Extraction of Domain Models (Code-to-Model Transformation): We realise this step by reverse engineering the highly cloned code patterns identified and semantically categorised in our code clone detection phase.
    \item Defining Meta Model from structural domains: Each structural domain is then characterized by an ontological basis, which describes the structural domains of the concepts and the relations between them to define a Meta Model for a DApp system (see \ref{fig:metamodel}). Besides this description of the structure of the domains, the meta model could also provide semantics for the concepts used. This meta model would capture the complete vocabulary of the transactions necessary to develop a smart contract based decentralised application (DApp).
\end{enumerate}

\begin{figure*}
\centerline{\includegraphics[width=0.9\textwidth]{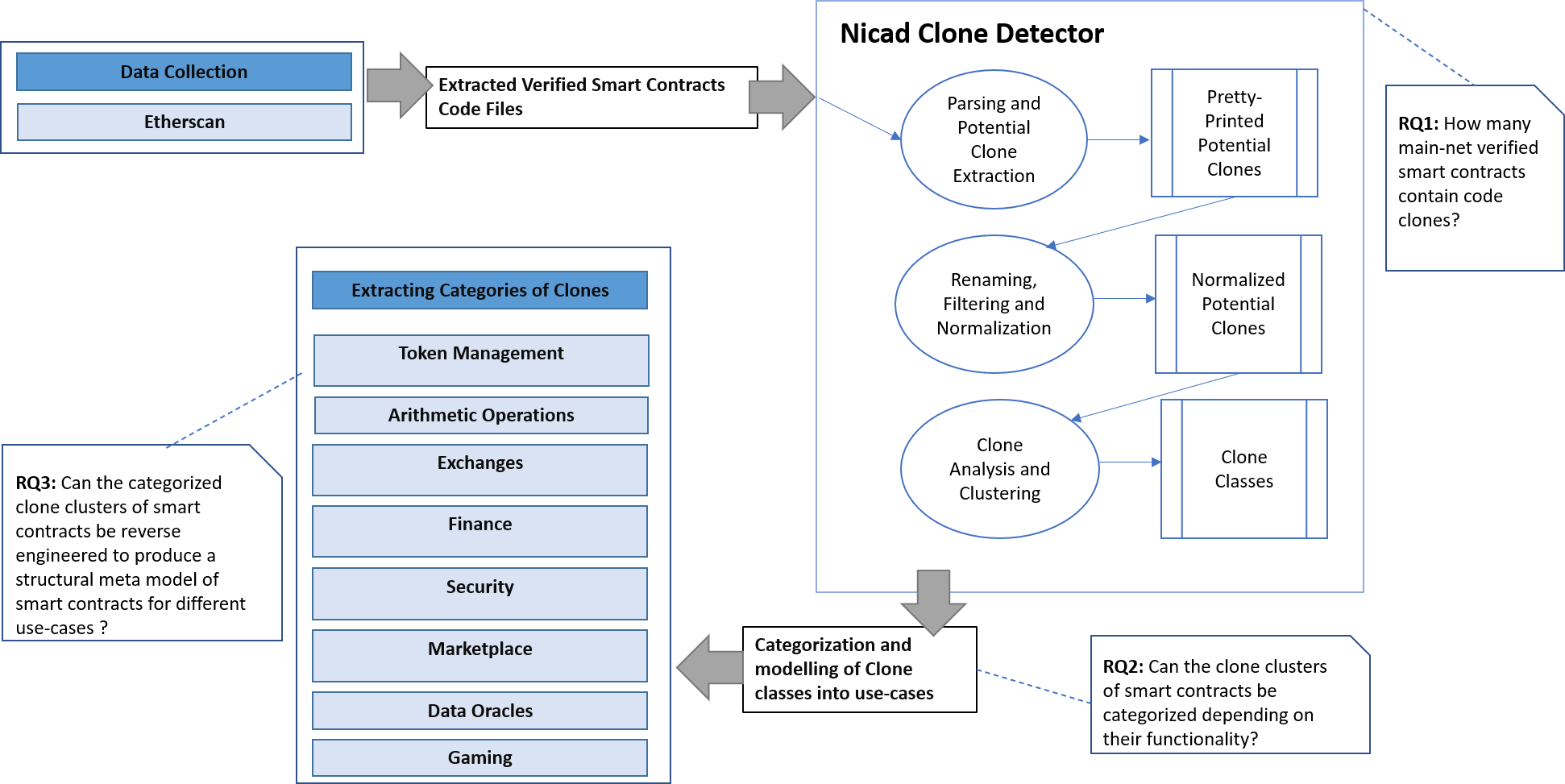}}
\caption{Overview of our approach at modelling domain categories from largest code clone clusters of Ethereum smart contracts}
%\vspace{-0.5 cm}
\label{fig:method}
\end{figure*}
\begin{table*}
\begin{center}
\caption{Classification Of Code Clones}
 \begin{tabular}{p{1cm} p{3.25cm} p{6cm}}
 \hline
\textbf{Clone Type} & 
\textbf{Clone Type Definition}& 
\textbf{Description}\\
\hline
1&Exact& Code clones of Type-1 are code fragments that are identical with variation in comments and blank spaces.\\
2&Renamed clone - Blind Renamed& Code clones of Type-2 are code fragments that are identical with variations in identifier names, literals, comments and blank spaces. Blind Renaming is the process of renaming all the identifiers to \textbf{X} without keeping a record of previously renamed identifier.  \\
2c&Renamed clone - Consistently Renamed& Code clones of Type-2c are similar to the Type 2 code clones except that they are consistently renamed, which is process of renaming the identifiers to \textbf{X}\textit{n}. Where, \textit{n} is the number of occurrence of the identifier with respect to the previously renamed identifier.\\
3-1&Near miss clone - Blind Renamed& Copied fragments with further modifications such as changed, added or removed statements, in addition to variations in identifiers, literals, types, white-space, layout
and comments.\\
3-2c&Near miss clone - Consistently Renamed& Code clones of Type-3-2c can be considered as a hybrid of the Type 3-1 and  Type 2c code clones in that they are similar code fragments with changed, added or removed statements, in addition to variations in identifiers, literals, types that are  consistently renamed. \\
4 & Functional Clone& Two or more code fragments that perform the same computation but are implemented by different syntactic variants.\\
\hline
\label{CloneClass}
\end{tabular}
\end{center}
\end{table*}
\section {Background}
\subsection{Ethereum Smart Contracts and Solidity}
Blockchain oriented software engineering to develop Decentralised applications (DApps) that execute smart contracts is complicated. Ethereum Smart Contracts \cite{Ethereum} are typically written in a high-level turing-complete programming language such as Solidity \cite{Solidity}, and then compiled to the Ethereum Virtual Machine (EVM) byte-code \cite{Ethereum}, a low-level stack-based language. These smart contracts cannot be updated once deployed to the main-net. This requires the developers to ensure that these smart contracts are attacks-resistant and reflect complete business logic desired by the DApp developers. A large portion of the problems with smart contract development could be pinpointed to the lack of formalization of blockchain oriented software engineering. The lack of standards and best practices makes smart contract development prone to problematic practices. 
%\textcolor{red}{Even though smart contracts cannot be changed once deployed to
%the blockchain, there is a method to develop upgradeable contracts.
%Ethereum provides a function named DelegateCall, which allows
%a contract to use code in other contracts, and all storage changes
%are made in the caller’s value. In this case, developers can develop
%two contracts, A and B. A is the proxy contract, which controls
%all the storage values, contract states, and Ethers. All the logic
%code is stored in contract B. Once errors are found, or new
%functionalities need to be added, contract B can be discarded, and
%contract A can call the code of the new contract. Based on this
%theory, OpenZeppelin, a famous smart contract organization, has
%provided a library to help developers develop upgradeable
%smart contracts in just a few lines.}
\subsection{Code-Cloning}
%Code clone detection \cite{NiCad} is an active research area and has been used in visualizing software evolution, maintaining software systems and even detecting malicious software. 
Reusing code for similar applications by cloning is an efficient way of minimizing the cost  and  time of software development. Research shows that the use of templates in software programming results in various advantages. It is especially very beneficial when there is a lack of knowledge base of a newly developed programming language. Developing templates from code clones and establishing them as standards or libraries \cite{EIP} also affects a  program by reducing its size making it easier to read and comprehend. 
%Various advantages of code clone detection and management are given \ref{CloneClass}.

\subsubsection{Types of Code Clones}
Code clones classification used in this analysis is given in Table[\ref{CloneClass}].

%ed as Type 1, 2 and 3 that correspond to the detection of Exact clones, Renamed or Parameterized clones, and Near-miss clones and Semantic c \cite{NiCad}. They are divided according to their speciality which 

\subsubsection{NiCad Code Clone Detector}
The NiCad clone detection tool is an efficient near-miss code clone detection tool \cite{NiCad}. In this paper, we demonstrate the detection of code cloning in Ethereum Smart Contracts using a static clone detection method. Our hypothesis is that near-miss clone detection will provide a means of implementing a model-driven development framework by producing a domain models of smart contracts for each of the popular domains in which smart contracts are being rapidly adopted. 
\subsection{Model-Driven Development for DApps}
%In the practice of software engineering, model-driven development is applied to lower the abstraction level of the system and to separate the components of a system. 
The model-centric or model-driven development approach focuses on utilization of models to describe the structure and the behavior of a system, which is then used to generate source code for a system. It can lower the complexity and create a better understanding of the system. \cite{modelSC}.
To automate and make more powerful maintainable smart contracts, we need to extend the MDE techniques to design and develop complex and decentralized applications. Smart contracts and DApps build on the premise of having a tremendous impact not only on the finance domain but also in other domains like gaming, supply-chain etc. However, its adoption requires novel skill sets that current software professional profiles fail to meet despite the increasing demand. Developing DApps using smart contracts would be easier if the learning curve for the needed skills would be more convenient. Model-driven software engineering can help in abstracting blockchain technology and, starting from these abstractions, enabling automated smart contract code generation. 

%The advantages of applying model-driven development to Decentralized Applications (DApps) development are listed below: 
%\begin{enumerate}
%    \item Higher level of abstraction:
%By hiding the technicalities of a smart contract, a programmer can focus on the domain logic to be developed.
%    \item Higher quality of the generated smart contract:
%Once modelled, those models can be verified, and thus developers can have confidence that the code generated will be correct and safe, thereby eliminating the need of regular auditing.
%    \item Higher inter-operability of the generated DApp:
%By using different code generators, a smart contract can be developed for different languages like Solidity, Vyper or Yul.
%\end{enumerate}

\section{Approach}
Our approach in producing domain models from existing smart contracts deployed on the Ethereum blockchain network is described in Figure \ref{fig:method}. It consists of following steps:
\subsection{Dataset}
Our data-set consisted of 10680 smart contracts that run behind a DApp. These smart contracts were collected from Etherscan \cite{Etherscsan}, a web portal that provides analytical data and repository of smart contracts deployed onto the Ethereum blockchain network. Etherscan provides verification of smart contracts and meta-data about a smart contract and its owner address. Partial meta-data for analysis of domain deduction was even collected from various literature surveys available online in the form of token white-papers, and GitHub repositories of contracts creators corresponding to a representative smart contract of a code clone cluster. 

\textbf{Demographics of the Analysed smart contracts}

To provide a proper context for the results, this section describes the demographics of the smart contracts in our dataset.
Our dataset of total 10680 sol files collected from Etherscan consisted of sub-contracts, libraries, interfaces, events, modifiers. To understand the models and the metamodel derived from our dataset, we provide the statistics of analysed smart contracts. As shown in the Table  \ref{demographics}, we analyzed 49,128 smart contracts which included main contract and its depended contracts. Other factors analysed included 8780 libraries, 2901 interfaces, 42101 events, 2901 modifiers. 
The extent of division of functionality into smaller sub-contracts is practised by software developers to achieve reduced complexity of contracts. Besides,
the events information shows that the execution information will be logged in Ethereum. The interfaces, libraries, modifiers and events information is used to derive structural models and templates. 

\begin{table*}
    \caption{Demographics of the Analysed 10680 Ethereum Smart Contracts}
\begin{center}
 \begin{tabular}{p{2.5cm}|p{2cm}|p{2cm}|p{2cm}| p{2cm}| p{2cm}|}
 \hline
\textbf{Total Sol Files}&\textbf{Contracts}& \textbf{Libraries}& \textbf{Interfaces}& \textbf{Events}& \textbf{Modifiers}\\
\hline
   \textbf{10680} & 49128 & 8780 & 2901& 42101 & 29014\\
\hline
\end{tabular}
\end{center}
\label{demographics}
\end{table*}

The final domain categorization of code clone clusters is provided in the following steps. 
\subsection{Using NiCad Near-Miss Code Clone Detection Tool}
The most recent version of NiCad code clone detector (v6.2) has built-in support for parsing and analysing source code files written in the Solidity programming language. However, the tool was not able to parse around 2800 smart contracts because of version incompatibility of the Solidity parser provided by the Nicad tool. We therefore updated the Solidity grammar to the latest version of the Solidity programming language to be able to parse and analyse all the smart contracts in our data-set. 
\subsection{Domain Categorization of Large Code Clone Clusters}
The main step of our research was to deduce domain categories based on the metrics of the code clone clusters produced by Nicad tool and even factoring in the semantics of code represented by these clusters. We leveraged the meta-data collected from various resources such as Etherscan \cite{Etherscsan}, StateofTheDApps \cite{SDApps} and DAppRadar \cite{DAppRadar} regarding the smart contracts in largest clusters to infer upon the domain categories. The final list of our domain categories was a result of an extensive manual browsing through the clusters to understand the semantics of the code fragments in the clusters and extracting relevant meta-data about the code fragments from online resources. 

\begin{table*}
\caption{Code Clones Found in a Dataset of 10680 Smart Contracts Collected From Etherscan \cite{Etherscsan}}
\begin{center}
\begin{tabular}{p{1.5cm}|p{1.25cm}|p{1.5cm}|p{1.25cm}|p{1.5cm}|p{1.25cm}|p{1.5cm}}
 \hline
&\multicolumn{2}{|p{2.5cm}|}{\textbf{Extractor Only}} & 
\multicolumn{2}{|p{2.75cm}|}{\textbf{Filtered and Blind-Renamed}}& \multicolumn{2}{|p{2.75cm}}{\textbf{Filtered and Consistent-Renamed}}\\
\hline
\textbf{Clone Type} & \textbf{Type 1}& \textbf{Type 3-1}&\textbf{Type 2}& \textbf{Type 3-2}& \textbf{Type 2c}& \textbf{Type 3-2c}\\
\hline
\textbf{Clone Pairs}& 17012	& 29645& 18534 &35115 & 18523&21487 \\
\hline
\textbf{Clone Classes}& 596& 607& 753 & 750& 752 & 762  \\
\hline
\textbf{Max diff threshold}& 0\%& 30\%& 0\%& 30\%&0\% &30\%  \\
\hline
\end{tabular}
\end{center}
\label{CloneDets}
\end{table*}

\begin{table*}
\caption{A Summary of the Categorization of the Top Clusters of Code Clones in a Dataset OF 10680 Ethereum Smart Contracts}
\begin{center}
 \begin{tabular}{p{0.5cm} p{2cm} p{2cm} p{2.5cm} p{2.5cm}}
 \hline
\textbf{S.no} & 
\textbf{Cluster Domain Category}& 
\textbf{Cluster Size}& 
\textbf{Accumulated Cluster Similarity Percentage(Min/Max)}&
\textbf{Domain Category deduced by clone detection type}\\
\hline
1 & Token Management & 547 & 90\%/100\%& Type 1,2,2c,3-1,3-2c  \\
2 & Arithmetic Operations & 309 & 91\%/100\%&  Type 1,2,2c,3-1,3-2c  \\
3 & Exchanges & 214 & 70\%/100\%& Type  2c, 3-1, 3-2c  \\
4 & Finance & 70 & 75\%/85\%& Type 3-2c    \\
5 & Data Oracles & 62 &70\%/77\%& Type 3-2c  \\
6 & Marketplace & 53 & 72\%/85\%&Type 3-2c   \\
7 & Gaming & 42& 77\%/98\% & Type 2c, 3-2c \\
8 & Security & 41 & 70\%/77\%& Type 2c, 3-2c   \\
\hline
\end{tabular}
\end{center}
\label{CloneCategorization}
\end{table*}
\subsection{Reverse Engineering of Domain Categories To Produce Domain Models}
After we formulated the list of domain categories exhibited by the largest code clone clusters in our analysis, we reverse engineered a representative smart contract of the cluster with an ultimate goal of producing structural models of the domain categories.
The reverse engineering step was performed using the \cite{Sol2Uml} tool available as an extension to the web-based Remix IDE for Solidity. 
%Our models depict code that can be re-used in our MDE framework (see Figure \ref{figure:initial})for developing Ethereum DApps with revised use-cases and increased diversity of functionality.
\subsection{Defining Meta Model}
\begin{figure*}
\begin{center}
\rotatebox[origin=c]{90}{\includegraphics[width=1.25\textwidth]{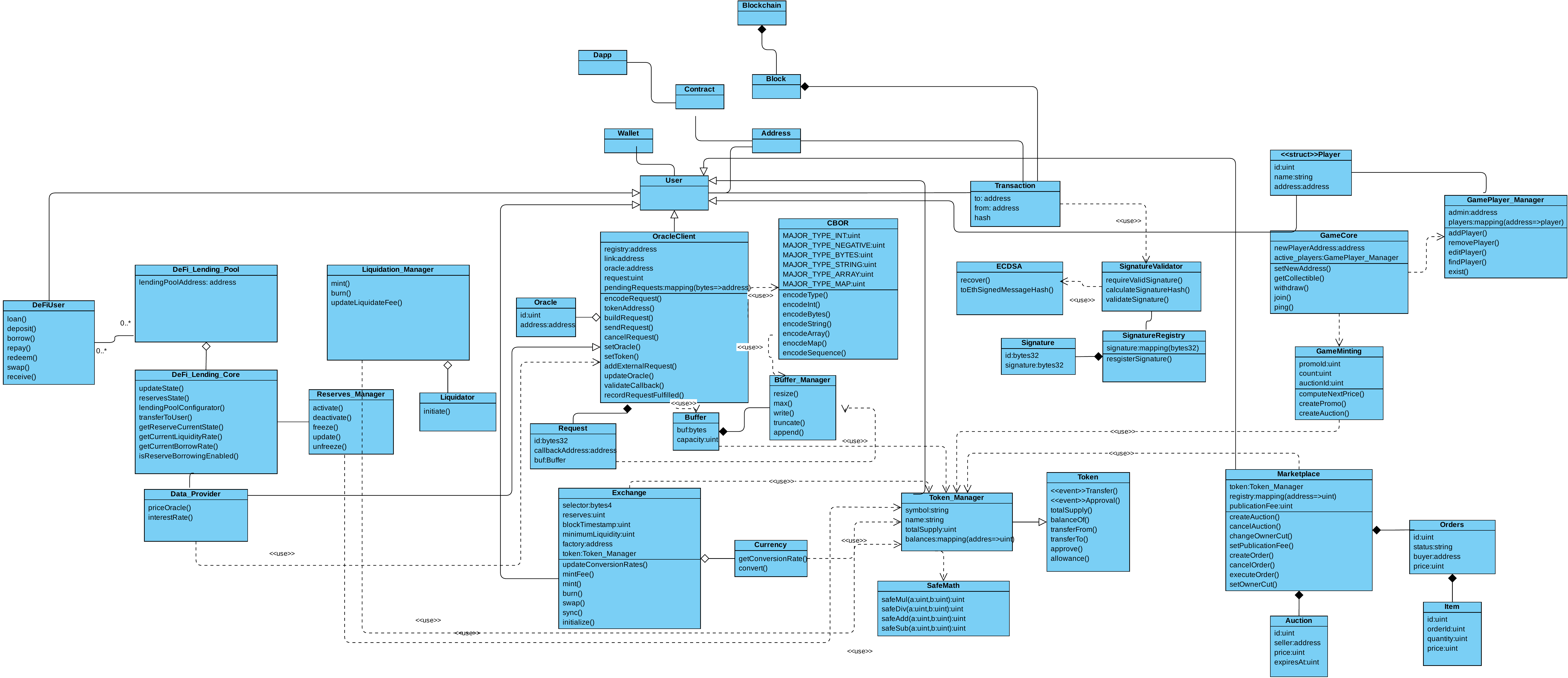}}
\caption{Meta Model for the mined domain models}
%\vspace{-0.5 cm}
\label{fig:metamodel}
\end{center}
\end{figure*}

The reverse engineered structural domain models showed certain overlap of patterns between domains. We therefore, aim at defining a Meta Model that represents common metadata required for deep semantic extraction of business rules and transformation of domain models.
The defined Meta Model describes the structural domain models of the concepts and the relations between them of a DApp system (see Figure \ref{fig:metamodel}). 
\section{Research Findings}
\subsection*{\textbf{RQ1:} Code Clones in Ethereum smart contracts - Analysis}

\textbf{Motivation.}
    First, we aim at identifying and characterizing the code clones in Ethereum smart contracts that are associated with DApps to be able to understand the diversity of use-cases implemented by Ethereum smart contracts. 
    
\textbf{Discussion.}
To answer our RQ1, we analyzed the output of NiCad near-miss clone detection tool. Some of the terms to understand before analysing the code clone detection report are: 
\begin{enumerate}
    \item Code Granularity: It refers to  the level of structural source unit taken into consideration while analysing a source code file for clones. In this paper, we configure NiCad clone detection tool to the \textit{functions} granularity to detect code clones in the functions of different Solidity smart contract code files. 
     \item Clone Pair: Two code fragments which are identical or similar are known to have a clone relation between them and are called a clone pair. 
      \item Clone Class: It is the maximum set of code fragments which contains a clone pair for each code fragment.
       \item Max. Difference Threshold: It is the percentage of difference in code fragments metric to be configured in the NiCad code clone detection tool to allow detection of code clones with a difference of up to the specified threshold. We configured it to allow code clones with a maximum difference threshold of up to 30\%. 
\end{enumerate}
The analysis results of a data-set of smart contracts are given in Table\ref{CloneDets}. We were able to detect code clones of Types 1, 2, 2c, 3-1, 3-2c using the NiCad near-miss code clone detection tool. A total of 40502 functions were extracted from the source code files and the clone size varied from 10 to 2500 lines of code. 

%\begin{figure}[t!]
% \subfloat[Token Management Model]{ \includegraphics[width=0.35\linewidth]{Token1.png}}\hspace{0.4cm}
%  \subfloat[Arithmetic Operations Model]{\includegraphics[width=0.48\linewidth]{SafeMath1.png}}
%  \caption{Token and SafeMath Libraries Model} \label{fig:TokenSafeMath}
%\end{figure}

\begin{figure}[t!]
\centerline{\includegraphics[width=0.75\textwidth]{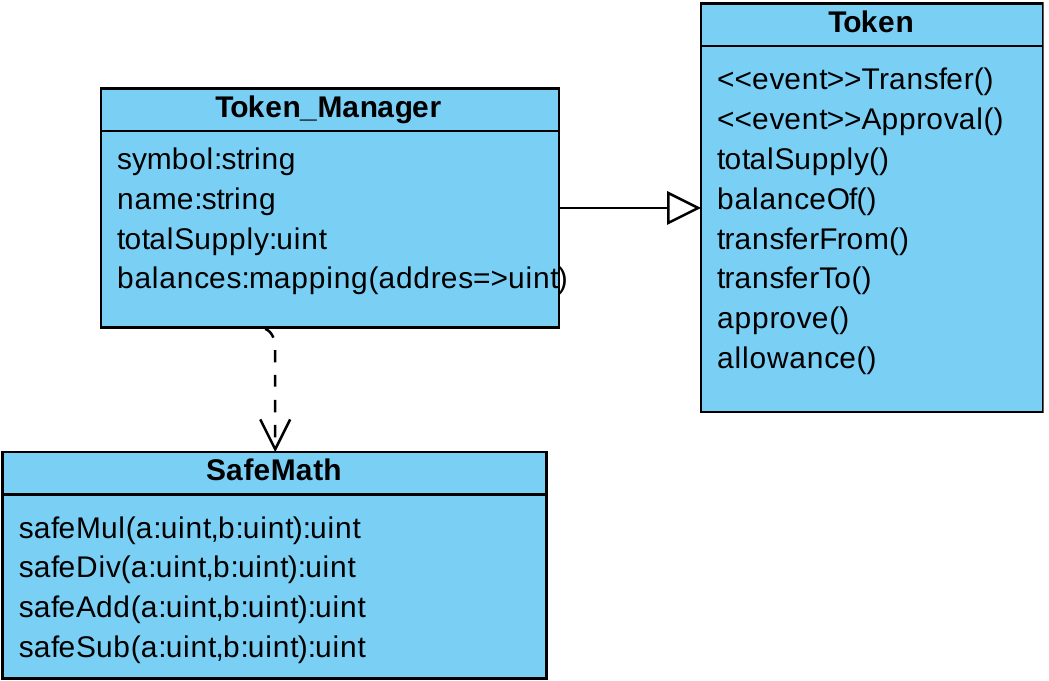}}
\caption{Token and SafeMath Libraries Model}
%\vspace{-0.5 cm}
\label{fig:TokenSafeMath}
\end{figure}

\textbf{Observation 1:} One of the observations we made while analysing code clone reports generated by NiCad is that even though there is a widespread adoption of Ethereum blockchain technology and smart contract, the source code files exhibit very similar functionalities. The majority of the smart contracts are either token contracts or token management programs. It can be inferred that the smart contract developers are mostly cloning code intentionally from the established standards available online (e.g., OpenZeppelin Libraries \cite{OpenZeppelin} to avoid any unforeseen security vulnerabilities that may arise because of the lack of knowledge and experience in this new technology. 
\begin{figure}[t!]
\centerline{\includegraphics[width=0.75\textwidth]{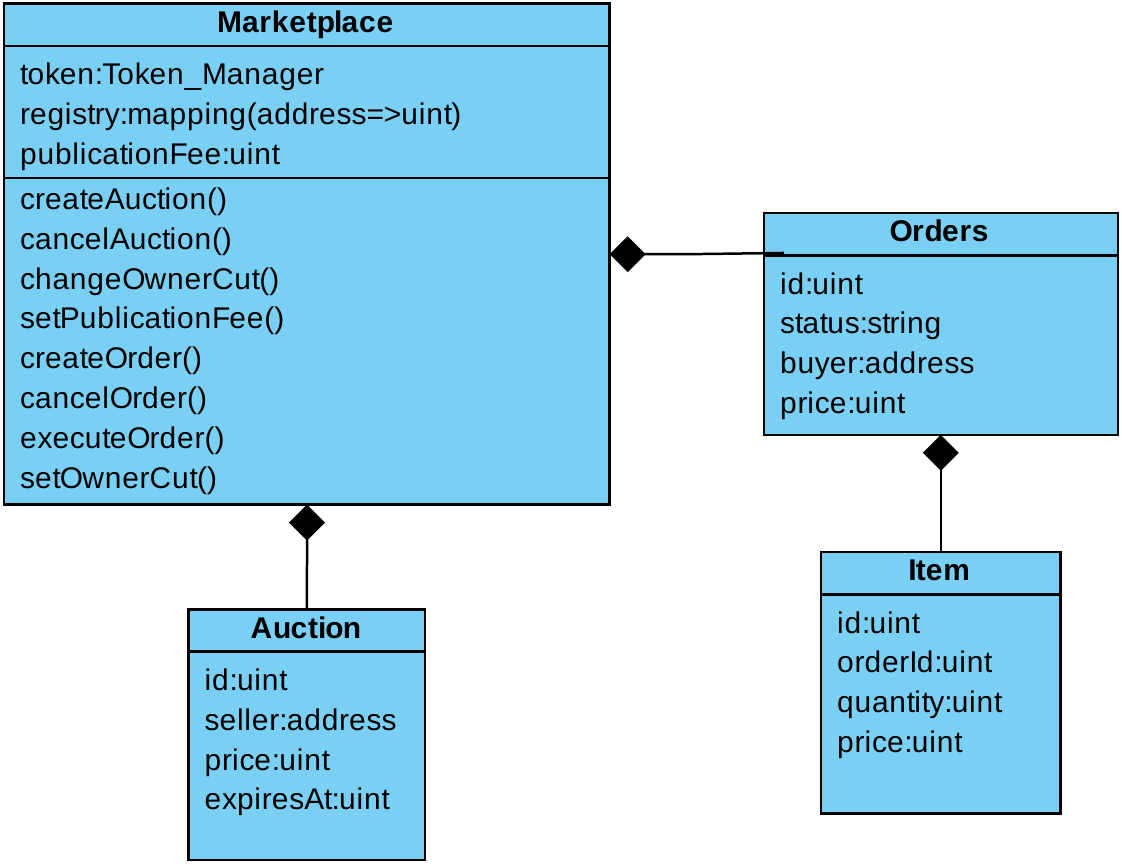}}
\caption{Marketplace Domain Category Model}
%\vspace{-0.5 cm}
\label{fig:Marketplace}
\end{figure}

\textbf{Observation 2:} Another noteworthy observation was that over 500 clone classes of smart contracts were code clone pairs arising from non-standard contracts.  We believe that this is because of the immutability characteristic of the blockchain technology that forces a smart contract developer to deploy the same contract again even with minor changes, such as updating the compatibility of their smart contracts to the latest version of the Solidity programming language. 

\subsection*{\textbf{RQ2:} Code Clones in Ethereum smart contracts -  Categorization}
 \textbf{Motivation.}
    To extract domain models for an ultimate goal of extracting business rules from existing Ethereum based decentralised applications (DApps), we categorize the largest code clone clusters identified in the code clone detection of Ethereum smart contracts. 

\textbf{Discussion.}
We approached the categorization of our code clone clusters depending on the cluster similarity percentages. Specifically, we categorised the code clone clusters into a domain category if a unique clone cluster has a similarity percentage of 70\% or higher. However, after a manual analysis of the clusters we came across multiple clusters of varying similarity percentages that could be categorised into a single domain category. To overcome this redundancy of clones and deduce one domain category of related functionalities, we analysed the source code of a representative smart contract in each of the clusters and accumulated the code clone clusters into a single domain category of related functionalities.
Our approach resulted in code clone clusters being categorised into following domain categories, [See Table\ref{CloneCategorization}]:
\begin{enumerate}
    \item Token Management: The largest code clone cluster was related to smart contracts that handled tokens. Most of the clusters were an implementation of the well-known token management standards such as ERC-20 \footnote{ERC: Ethereum Request for Comments, Common ERC standards define a required set of functions for a token type \cite{EIP}} (Token management standard for fungible tokens) and ERC-721 (Token management standard for non-fungible tokens). Almost all the other domain categories deduced by our analysis leveraged this cluster of code clones to be implemented along with their specific functionalities. 
    \item Arithmetic Operations: Earlier versions of the Solidity programming language was detected with certain arithmetic vulnerabilities arising from it syntax. Before the release of the latest version of the Solidity programming language, the smart contract developers adopted the practice of implementing the \textit{SafeMath} library provided online to safely perform the arithmetic operations in a smart contract. This practice is evident from our analysis and the second largest cluster of code clones contributed to the implementation of \textbf{SafeMath} library. Although the latest version of the Solidity programming language has this vulnerability resolved, our data-set consisted of smart contracts ranging from the  versions 0.5.0 to 0.8.0. \cite{Solidity}
    \item Exchanges: The exchanges domain category consisted of use-cases involving trading of digital assets built upon the Ethereum blockchain technology. The functionalities consisted but not limited to minting, burning and swapping of digital assets like fungible and non-fungible tokens.  
    
    \textbf{Observation 3:} We observed that the popular cryptocurrency exchange names like  \cite{CoinBase},   \cite{Kraken}, \cite{cryptoWatch} were listed in this cluster. One of the surprising notes was that some of the above mentioned established and popular exchange platforms had a similarity percentage of up to 100\% in this cluster. This information could be used to infer upon the extent of plagiarism or inadequate copyright policies in the world of blockchain technology based decentralised applications. 
    
    \textbf{Observation 4:} 60\% of the smart contracts in this cluster also leveraged the \textit{Data Oracles} code clone cluster discussed later in this section. It was also one of the clusters that had smart contracts leveraging \textit{Token Management} code clones extensively.  
    \item Finance: One of the industries that has the most potential of getting revolutionized by the Ethereum blockchain technology is the Finance industry. DeFi (Decentralised Finance) refers to financial services like accessing and handling financial assets in a universal and decentralized manner using the blockchain technology.
    
    \textbf{Observation 5:} Despite all the hype around the DeFi solutions provided by the Ethereum blockchain technology and smart contracts, our code clone analysis showed that the smart contracts developed behind the DeFi DApps were very limited and homogeneous when compared to the use-cases provided by traditional financial applications. The \textit{DeFiLendingPool} smart contract had a similarity percentage of upto 71\% that consisted of use-cases revolving around lending, borrowing, withdrawing and depositing of assets. It leveraged a form of \textit{Token Management} cluster of code clones as well.
   
%\begin{figure}[t!]
% \subfloat[Exchange Domain Category Model]{ \includegraphics[width=0.35\linewidth]{Exchange.png}}\hspace{0.4cm}
 % \subfloat[Decentralized Finance Domain Category Model]{\includegraphics[width=0.48\linewidth]{DeFi1.png}}
 % \caption{Exchange and Decentralized Finance Domain Model} \label{fig:ExchangeDeFi}
%\end{figure}

\begin{figure}[t!]
\centerline{\includegraphics[width=.75\linewidth]{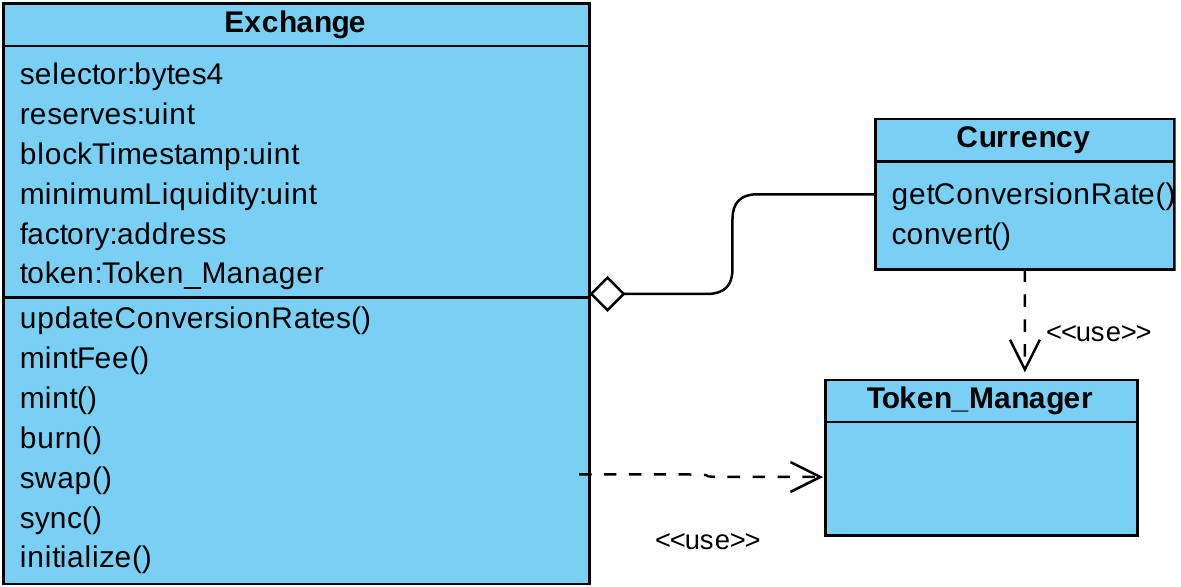}}
\caption{Exchange Domain Category Model}
%\vspace{-0.5 cm}
\label{fig:Exchange}
\end{figure}

\begin{figure*}
\centerline{\includegraphics[width=1\linewidth]{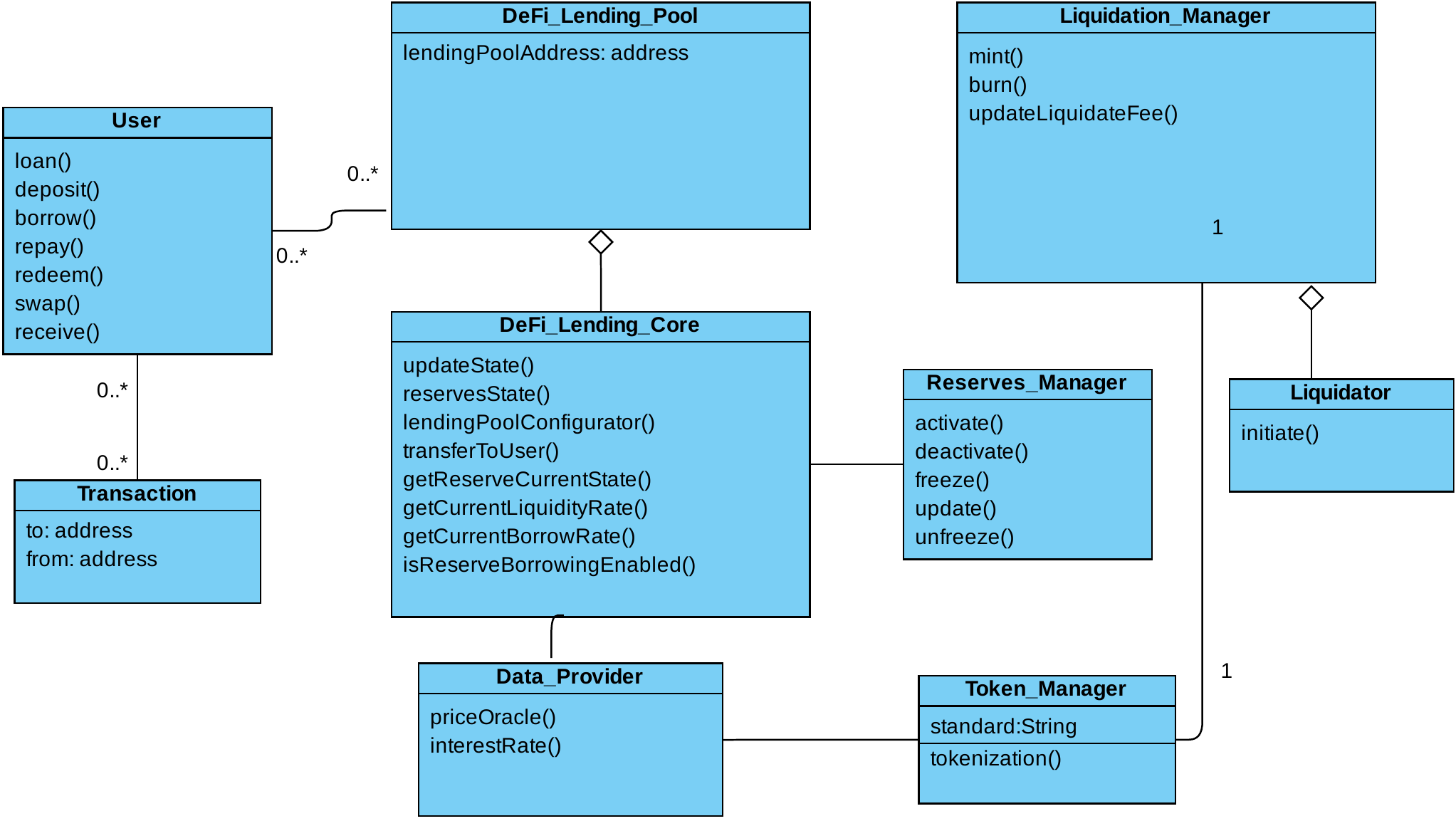}}
\caption{Decentralized Finance Domain Category Model}
%\vspace{-0.5 cm}
\label{fig:DeFi}
\end{figure*}

\begin{figure}[t!]
\centerline{\includegraphics[width=0.75\linewidth]{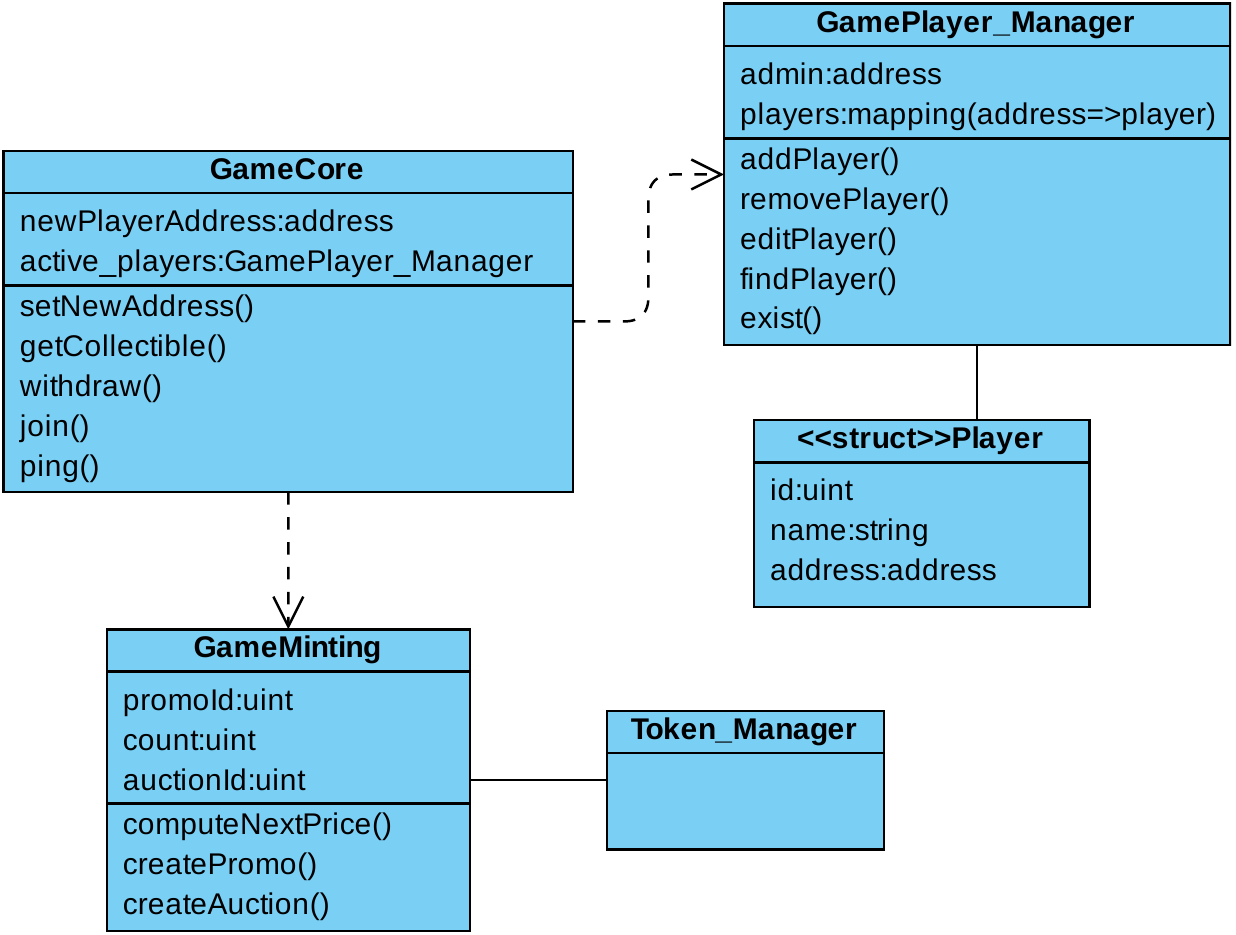}}
\caption{Gaming Domain Category Model}
%\vspace{-0.5 cm}
\label{fig:game}
\end{figure}
\begin{figure*}
\centerline{\includegraphics[width=0.85\linewidth]{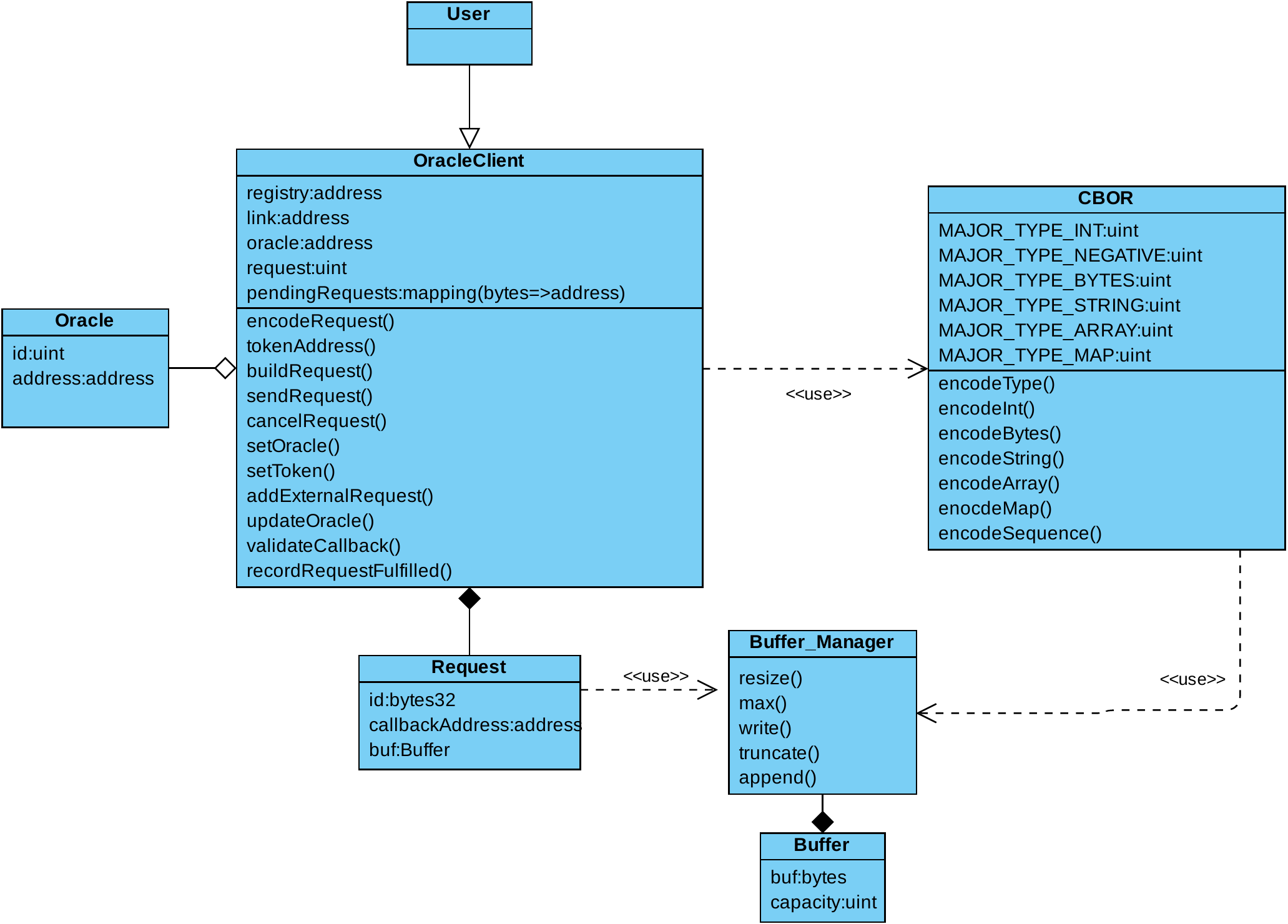}}
\caption{Data Oracle Domain Category Mode}
%\vspace{-0.5 cm}
\label{fig:DataOracle}
\end{figure*}
%\begin{figure*}[t!]
% \subfloat[Oracle abstract contract]{ \includegraphics[width=0.5\textwidth]{Oracle.png}}\hspace*{\fill}
%  \subfloat[Oraclize abstract contract]{\includegraphics[width=0.5\textwidth]{Oraclize.png}}
%  \caption{Data Oracle Domain Category Model} \label{fig:DataOracle}
%\end{figure*}

 \begin{figure}[t!]
\centerline{\includegraphics[width=0.75\textwidth]{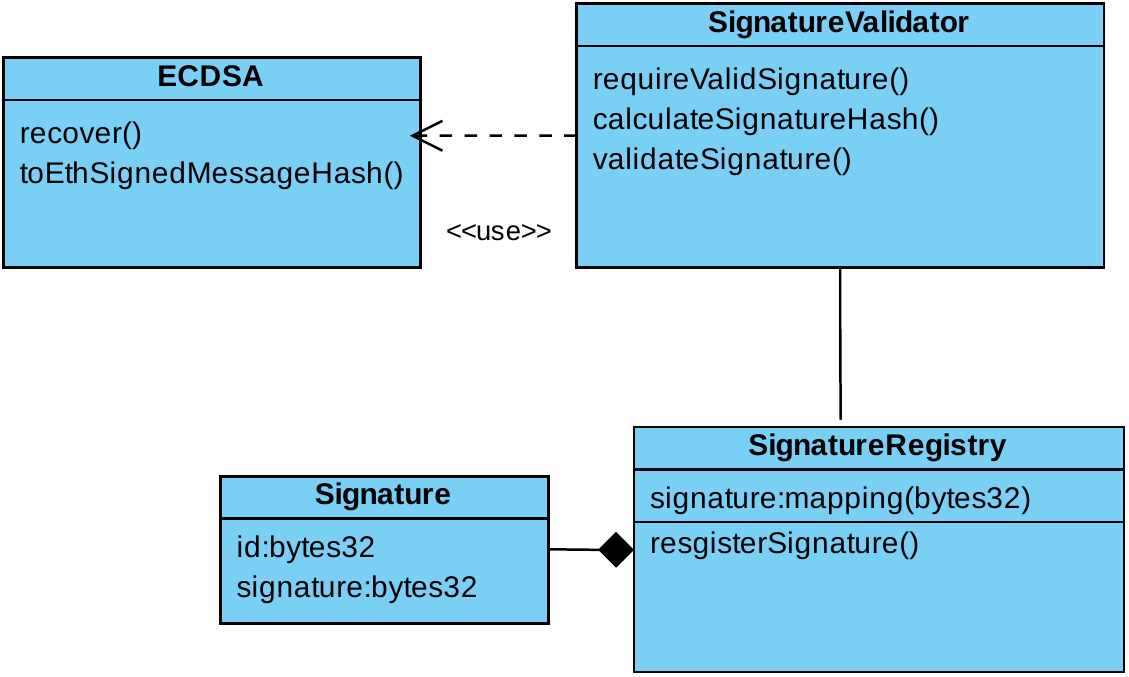}}
\caption{Elliptical Curve Digital Signature Algorithm (ECDSA) - Security Domain Category Model}
%\vspace{-0.5 cm}
\label{fig:security}
\end{figure}
    \item Security: Blockchain technology is based on cryptographic signatures to authenticate an owner of an address without disclosing its private key. These signatures are primarily used for signing transactions and sometimes also to sign arbitrary messages. Ethereum blockchain technology uses the Elliptic Curve Digital Signature Algorithm \cite{ECDSA}. ECDSA is only a signature algorithm that cannot be used for encryption and consists of two integers: \textit{r} and \textit{s}. Ethereum uses a recovery identifier \textit{v}, thereby making the signature as \textit{\{r, s, v\}}. These message signatures can be used to verify the ECDSA signatures using smart contracts. A function to recover the address of the private key of the message signatory uses a pre-compiled contract at address \textit{0x1}. Signature verification in smart contracts finds many useful applications such as, multi-sig contracts and exchanges. The code clone cluster representing the ECDSA message signatures validation functionalities implemented a form of standard ERC-1271 provided by the Ethereum blockchain technology.
    \item Marketplace: Blockchain technology has encompassed \$250 million market of NFTs (non-fungible tokens) that are traded as digital assets. NFTs include artefacts ranging from digital trading of art to virtual real estate and gaming collectibles. A fungible token (FT) can be directly traded with another FT, however, same cannot be said about NFTs. These marketplace DApps that allow trading of NFTs ranged anywhere from \$10 to hundreds of thousands of dollars and yet lack security and proper copyright validation checks. 
    
    \textbf{Observation 6:} While analyzing the code clone clusters representing trading of NFTs in smart contracts, we made an interesting observation that the domain categories are not clearly separated in the Ethereum DApps ecosystem. For example, the DApps \textit{AxieInfinity} \cite{AxieW} and \textit{DecentraLand} \cite{DecentralandW} contained code fragments from Marketplace domain category and they also exhibited strong code functionalities categorised under \textit{Gaming} domain category. 

    \item Data Oracles: The blockchain technology is a somewhat a closed network with complexity revolving around extracting data from the blockchain network and access data from external sources into the blockchain network. However, with blockchain based DApps receiving growing interest from different industrial domains, it is necessary to enable a smart contract to access external data required to control the execution of the business logic. Data oracles provide an efficient interface to access external data from various data oracles in to a smart contract. These oracles can be modified to cater for different types of data depending on the industry and requirements. \textit{Oraclize or Provable Things} \cite{Provable} and \textit{ChainLink} \cite{ChainLinkW} are the most widely used data oracles in the blockchain network. A data request to these data oracles is validated against the inclusion of the following information from the blockchain network: a data source type, a query, and an optional authenticity proof. 
    \item Gaming: Gaming DApps in blockchain networks have gained popularity because of the involvement of trading of financial/digital assets in the form of NFTs and sometimes even FTs while playing the game. 
    
    \textbf{Observation 7:} We observed clusters of code clones depicting a similarity percentage of 100\% of successful and popular Ethereum based gaming DApps such as Cryptokitties  \cite{cryptokitties} and AxieInfinity \cite{AxieW}. 
    Given the fact that there are not many strict copyright protection systems in place for blockchain DApps, there is an evident plagiarism of gaming DApps by the DApps developers to benefit financially.
  
\end{enumerate}
\subsection*{\textbf{RQ3:} Code Clones in Ethereum smart contracts - Modelling}
\textbf{Motivation.}
    To be able to re-use the existing models in the Ethereum smart contracts and modernize them to integrate newer business rules to diversify the use-cases of Ethereum based DApps, we reverse engineer the extracted code clones from Ethereum smart contracts to produce domain models. 

\textbf{Discussion.}
Figures \{\ref{fig:TokenSafeMath}, \ref{fig:Marketplace}, \ref{fig:Exchange},\ref{fig:DeFi}, \ref{fig:game}, \ref{fig:DataOracle}, \ref{fig:security}\} show the structural models produced by a manual reverse engineering of code fragments from the largest code clone clusters of smart contracts categorised into domains in the above step. 

The \textbf{Token Management} domain category mainly represents the ERC20 and ERC721 \cite{EIP} token standards and therefore the domain model extracted from reverse engineering the code clone cluster representing the functionalities mostly cloned by smart contracts is showed in Figure \ref{fig:TokenSafeMath}. This model lists functions related to management of tokens in a DApp, such as transfer, checking balance, setting allowance of transfer, approval of transfer etc. 

Similarly the \textbf{Arithmetic Operations} domain category uses the \textit{SafeMath} (see Figure\ref{fig:TokenSafeMath}) interface to represent the operations that can be performed with the surety of protecting the smart contract from the vulnerability that existed in the Solidity programming language regarding \textit{Integer Overflow}. 

\vspace{-0.05 cm}
The \textbf{Exchanges} domain category model (see Figure\ref{fig:Exchange}) represents the functions that were cloned the most by the currently popular cryptocurrency exchange platforms available. The functionalities represented in this model deal with minting, burning and swapping of digital assets and currencies.

\textbf{DeFi} domain category model (see Figure\ref{fig:DeFi})was reverse engineered from code clone clusters that required more semantic analysis of the code than the rest of the domain categories. The \textit{DeFiLendingPool} smart contract represented most of the functions that were cloned by other DeFi applications, which consisted of use-cases related to lending, borrowing, depositing and withdrawing of digital assets from a smart contract defined lending pool. 

\textbf{Data Oracles} domain category uses the \textit{Provable} data oracle available online to access external data into a smart contract for efficiently executing a business logic. 

The domain model depicts the \textit{Oracle} and \textit{Oraclize} \textit{Abstract} contracts that easily integrate the external data into a blockchain application. (see Figure\ref{fig:DataOracle})

Trading of digital assets, primarily NFTs, is realised by the \textbf{Marketplace} domain category. The functions related to the business logic of this domain consisted of creating and cancelling of auctions, maintaining order history, order management such as receiving, executing, filling and cancelling a order. It also leveraged the \textit{Token Management} standard of ERC721 to manage the exchanges of NFTs. (see Figure\ref{fig:Marketplace})

The \textbf{Gaming} domain model represents the largest code clone clusters related to a smart contract behind a Ethereum based gaming DApp. However, the model extracted from this cluster is generic and does not relate to the business rules of the game specifically. The model shows the structure of smart contracts to manage ownership of the game, mint new tokens and get new collectibles in the game. (see Figure\ref{fig:game})

The \textbf{Security} model represents the signature validation structure implemented by smart contracts using the ECDSA algorithm to sign the transfers and abitrary messages. (see Figure\ref{fig:security})

\section{Related Work}
\textbf{Code Cloning in Ethereum Smart Contracts.}
In the space of code cloning analysis and detection tools for Ethereum smart contracts, there are only a few studies so far. The most similar study to ours is the work of Kondo et al. \cite{codecase} who study code cloning in Ethereum at the source code level with a goal to quantify the number of clones in Ethereum and their characteristics. They follow the ideology of positive impact of code cloning in software systems to avoid taking the risk of writing vulnerable code. The authors compare the code clone clusters to an online library of publicly available open-source smart contracts on OpenZeppelin \cite{OpenZeppelin}. The analyse the code cloning in a data-set of 33,073 smart contracts and resulted in a clone ratio of around 80\%. \cite{codecase} employs a tree-based clone detector, Deckard \cite{Deckard} at the source code level and analyse the detected code clones to understand the Ethereum blockchain technology ecosystem. 
We agree with the authors observation that although the blockchain technology is gaining adopting in a wide range of use cases, currently there is a lack of diversity in its applications.

Another research study by Gao et al. produced a code clone detection tool for Ethereum Smart Contracts called SmartEmbed. This tool identifies code clones from the source code of smart contracts using the code embedding and similarity checking techniques. SmartEmbed is aimed at quantifying clones in Ethereum Smart Contracts and identify the clone-related vulnerabilities. SmartEmbed was evaluated on a data-set of around 22000 smart contracts contracts and resulted that the clone ratio in Ethereum Smart Contracts is close to 90\%.   \cite{SmartEmbed}

Other than the above mentioned two studies, some researchers also analysed the code cloning in Ethereum Smart Contracts at the byte-code level. 

A research study is by Liu et al. that detects code clones in Ethereum Smart Contracts using the byte-code generated by the Ethereum Virtual Machine (EVM) of each deployed smart contract. 
is more challenging and less advantageous towards derivation of models by reverse engineering of the code clone as the byte-code factors in the version number of the Solidity programming language being used and could potentially miss some of the core functional code clones across smart contracts. \cite{LiuByteCode}

We expand upon these papers by investigating code clones of type 3, which focuses on near miss renamed code clones rather than just the code clones resulting from identical code reuse. Moreover, we contribute to the research on code re-use in Ethereum smart contracts by providing a categorization of these investigated large clusters of code clones. These clusters are further subjected to reverse engineering to produce domain models to facilitate code reuse at the model level. 

\textbf{Model-Dirven Development and Ethereum based decentralised applications (DApps) development.}
A Model-Driven Reverse Engineering (MDRE) approach proposed in \cite{cosentino} aims at extracting business rules out of a Java application. To achieve this goal, the initial step is the model discovery from Java code followed by variable classification and domain variables model creation. The final step of their approach is to extract the business rules model from the domain variables model. 

An extensive research on application of model-driven development approach to smart contracts in general was produced in \cite{Boogaard2018AMA}. In this research study, the author conducted a literature survey to gain the requirements and techniques to formulate a model-driven development method for smart contracts development. The proposed model-driven method is evaluated using a standard case study and an experiment involving smart contract developers. However, this research study does not support the frequently evolving programming language requirements and established standards to the smart contracts development. 

An initial step to applying model-driven engineering to generate smart contracts for IoT and cyber-physical systems by Garamvolgyi et al. talks coordinating the usage of aforementioned system elements from behavioral models, specifically UML statecharts.\cite{modelSC} 

A DEMO, BPMN and UML based approach by Skotnica et al. proposes a visual domain-specific language called DasContract. This approach creates models that contain all possible execution paths according to the DEMO transaction axiom and considers modeling of smart contracts using three subsystems - data, processes and forms.  \cite{bachelorThesis}

An agent-based approach by Frantz et al. is a computational modeling paradigm in which phenomena are modeled as dynamical systems of interacting agents.\cite{agentBasedApproach}

An MDE tool, Lorikeet by Weber et al.
\cite{Lorikeet2018Weber} uses BPMN models and fungible/non-fungible registry data schema along with the bpmn-js modeling library to develop a business process execution and asset management focused approach for developing smart contracts. 
Another approach lead by Weber,
\cite{Caterpillar2017Weber} uses Camunda which is an open-source workflow and decision automation platform to model business processes to be executed using smart contracts. 

The behavioral modeling of smart contracts is realised by Mavridou et al.
\cite{FSolidM2018}, which proposes using finite state machine based approach to model smart contracts execution. This tool is built upon WebGME MDE tool and provides a graphical editor along with automatic code generation with security plugins. 

A recent study by Hamdaqa et al.
\cite{SAMMDE} propose a feature-oriented domain analysis technique, \textit{iContractML}, to extract the a reference model based on the similarities and differences between the three blockchain platforms namely, IBM Hyperledger Composer, Azure Blockchain Workbench, and Ethereum. This research further realizes the derived reference model into an MDE framework to enable smart contract developers in modeling and generating smart contracts that are compatible with multiple blockchain platforms. 
Our approach differs from \textit{iContractML} extensively as the technique used to extract domain models in our approach uses reverse engineering of frequently used code patterns, whereas, \textit{iContractML} uses a forward engineering approach for domain analysis. The motivation behind \textit{iContractML} is to promote scalability across different blockchain platforms and our approach focuses on realizing the most frequently used code patterns by smart contract developers in the form of a library of models, thereby, enabling secure reuse of code libraries and better comprehension of smart contract functionalities. 

\section{Threats to Validity}
\subsection*{\textbf{Threats to Internal Validity}}
One of the factors that can affect our results is that our approach is a code clone based domain model mining method which can only derive instances of repetitively used domain models in smart contracts. Though our method can account for variations in models, this cannot be relied on for extending the current state of usability of smart contracts. To overcome this limitation of allowing only a threshold of variation in domain models during the code clone detection method, we analysed and updated the derived domain models depending on their applicability in a specific business process. We recognize that further research in the area of business process modelling for Ethereum smart contracts may benefit our results.  

\subsection*{\textbf{Threats to External Validity}}
The dataset analyzed in this research consisted of a subset of all the Solidity smart contracts deployed onto the Ethereum main-net that corresponds to DApps.  Our dataset does not include the smart contracts that do not correspond to a DApp deployed onto the Ethereum main-net. Moreover, DApps are being built on top of other blockchain networks as well and might represent different characteristics when compared to the smart contracts analyzed in this research. We recognize that further research with a larger dataset may strengthen the correlation established in our results.  We highlight that the contribution of this paper is to conduct an initial study towards producing an MDE framework to efficiently develop DApps that are maintainable and secure. However, it should be considered that our research results may not be applicable to all the DApps built on top of other blockchain networks as well.

\section{Conclusion and Future Work}
The main reason behind code cloning in Ethereum Smart Contracts is the stigma around the complexity involved behind securing a smart contract and using the features provided by the Solidity programming language and the Ethereum blockchain technology to efficiently develop a decentralised application (DApp). The lack of experience, resources and an active involvement of online community in developing DApps have made the software developers of traditional programming languages to rely on standards and existing deployed DApps available, thereby making the Ethereum smart contracts ecosystem very homogeneous in functionality. We conducted this exploratory analysis of code clones in Ethereum Smart Contracts as an initial step towards developing a model-driven development framework for developing smart contracts based DApps. We perform code clone detection of Type 1,2, 2c, 3-1, and 3-2c using the NiCad code clone detection tool. We especially leverage the the near-miss code clone detection of Type 3-1, 3-2c provided by NiCad to understand the functionalities implemented by various smart contracts of a similar background or domain to semantically categorise the highly cloned code clusters.

The future work in this research would be to perform an analysis of use-cases implemented by the domains listed in our categorization of code clones, to be factored in the structural domain models of smart contracts through M2M transformations for each of the listed domain categories. Eventually we aim at producing a model-driven development framework for fast and efficient development of DApps with increased maintainability, extensibility and diversity of business use-cases. 
%\begin{acknowledgements}
%\end{acknowledgements}

\clearpage
\bibliographystyle{IEEEtran}
\bibliography{CodeClonesBib.bib}
\nocite* 
\end{document}